# GAC: Energy-Efficient Hybrid GPS-Accelerometer-Compass GSM Localization

Moustafa Youssef
Computer and Systems
Engineering Department
Faculty of Engineering
Alexandria University
Email: moustafa@alex.edu.eg

Mohamed Amir Yosef
Computer and Systems
Engineering Department
Faculty of Engineering
Alexandria University
Email: mamir@alex.edu.eg

Mohamed El-Derini
Computer and Systems
Engineering Department
Faculty of Engineering
Alexandria University
Email: elderini@ieee.org

*Abstract*—Adding location to the available information enables a new category of applications. With the constrained battery on cell phones, energy-efficient localization becomes an important challenge. In this paper we introduce a low-energy calibration-free localization scheme based on the available internal sensors in many of today's phones. We start by energy profiling the different sensors that can be used for localization. Based on that, we propose GAC: a hybrid GPS/accelerometer/compass scheme that depends mainly on using the low-energy accelerometer and compass sensors and uses the GPS infrequently for synchronization. We implemented our system on Android-enabled cell phones and evaluated it in both highways and intra-city driving environments. Our results show that the proposed hybrid scheme has an exponential saving in energy, with a linear loss in accuracy compared to the GPS accuracy. We also evaluate the effect of the different parameters on the energy-accuracy tradeoff.

## I. INTRODUCTION

The number of location-based applications has been growing exponentially in the last few years. With the increase in the number of mobile phones that have location capabilities, many location-aware applications have emerged. Examples include Loopt [1], Google's Latitude [2], Micro-Blog [3], MetroSense [4], Place-Its [5], PeopleNet [6], MyExperience [7], among others. The energy-constrained nature of mobile phones has made energy-efficient localization a hot research area.

A number of localization technologies have been proposed for cellular phones. The Global Positioning System (GPS) [8] is the most commonly used one and the most ubiquitous. However, one of the main drawbacks of GPS is that it is an energy-hungry device that can drain the battery of the phone in a few hours [3]. Different approaches based on time, angle, and received signal strength (RSS) have been proposed [9] but were not deployed due to the limitation of technology and APIs at that time. Only recently have we witnessed the implementation of localization systems for cellular phones. RSS-based GSM systems [10], [11] and city-wide WiFi-based localization systems [10], [12] have been proposed but they either suffer from low accuracy or high energy consumption [3].

With the advances in cell phone hardware, new internal sensors have been introduced including the accelerometer and compass. An accelerometer is a device that measures the acceleration the device experiences relative to free-fall. The compass measures the magnetic field strength applied to the three axes of the compass. Consequently, it can give an estimate of the angle between the device and the north direction. Many of today's popular cell phones come with a 3-axis accelerometer and compass. Accelerometers have been used to detect the device orientation, e.g. to auto-rotate the screen. Many applications, e.g. games, also utilize the accelerometer to detect tilting of the mobile. Recent research [13]–[15] has suggested using the accelerometer and compass for localization, mainly based on detecting the ***human activity recognition***. For example, in [13], [14] accelerometers have been used to detect whether the user is sitting, standing, or walking. Based on this activity recognition, the localization system can be used to differentiate between the user sitting in a coffee-shop or walking in an adjacent store. Similarly, the CompAcc system [15] uses accelerometers to detect the number of human steps, based on the human ***movement pattern***, to obtain the distance traveled and matches the direction obtained from the compass with the road direction using a map of the area of interest. In this paper, we take a different approach for localization using the accelerometer and compass based on ***Newton's laws of motion***. The basic idea is to combine the compass and accelerometer readings to estimate the mobile location using an initial position and velocity. The compass determines the direction of motion while the accelerometer reading can be integrated twice to determine the displacement of the phone along the direction of motion. Due to the noise in the accelerometer and compass readings, error accumulates as time goes on. Therefore, we turn on the GPS briefly and infrequently to obtain an accurate position estimate. Although the idea of using the accelerometer and compass has been known for a long time in Inertial Navigation Systems (INS), adopting it to mobile devices introduces many challenges among which is the fact that sensors available in the mobile are cheap inaccurate sensors. Moreover, the whole system is vulnerable to shaking and minor changes in orientation. Most importantly, energy consumption is not a constrain in traditional INS that are usually connected to non-constrained sources of energy.

We start by profiling the energy consumption of the different sensors on the HTC Dream Android-based[1] cell phones to quantify the advantage of using the accelerometer/compass as compared to the GPS. We then describe our hybrid GPS-accelerometer-compass (GAC) scheme in details along with its implementation and discuss how our scheme handles shaking and minor changes in orientation along with reducing noise in sensors measurements. Evaluating our system under two different testbeds, representing highways and intra-city driving conditions, shows that the hybrid scheme leads to an exponential saving in energy consumption compared to using GPS alone with a linear decrease in accuracy. We also study the effect of changing the system parameters on the accuracy-energy saving tradeoff.

The rest of the paper is organized as follows: Section II presents the energy profiling experiment. Section III gives the details our GAC system. Section IV evaluates the GAC system and compares it to other schemes in terms of power consumption and accuracy. Finally, Section V concludes the paper and gives directions for future work.

## II. POWER PROFILING FOR INTERNAL SENSORS

In this section, we present our experiment to profile the power consumption of the different sensors that can be used for localization available on the HTC Dream Android-based phone namely: the GPS, accelerometer, and compass. We start by describing our experimental setup followed by the profiling results.

### A. Experiment Setup

The experiment setup is divided into two parts: hardware and software.

*1) Hardware Setup:* In order to measure the power consumed by the Android device, the following setup is used. The Android battery is removed and replaced with a fake battery connected to a Monsoon power monitoring device [16]. The Monsoon device supplies the phone with the required power and measures both the instantaneous current and voltage used by the phone. Since the voltage is constant, the current can be used as an indication of power consumption. The Monsoon device is connected to the PC and reports all monitored values through a USB connection. The hardware setup is shown in Figure 1.

*2) Software Setup:* There are two parts of the software: an Android application running on the Android device and a Monsoon client running a Windows machine. The Android application turns ON and OFF different sensors of the Android device, e.g. GPS, accelerometer, and/or compass. The Monsoon client reports all measured values by the Monsoon hardware and logs them in both graphical and textual formats. The text files are analyzed offline to get an estimate for the power consumption of the different sensors of the mobile. This is achieved by comparing the average power consumed in the standby state with that consumed when the sensor of interest is running.

[1]Android is a new *open source* operating system for mobile phones that is rapidly gaining market share.

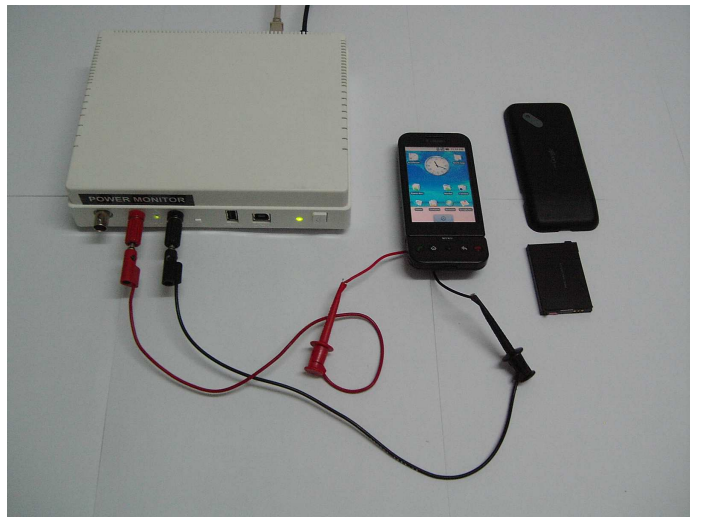

Fig. 1. The hardware setup for the power profiling experiment. The HTC Dream's battery is removed (right) and replaced by a fake battery connected to the Monsoon device.

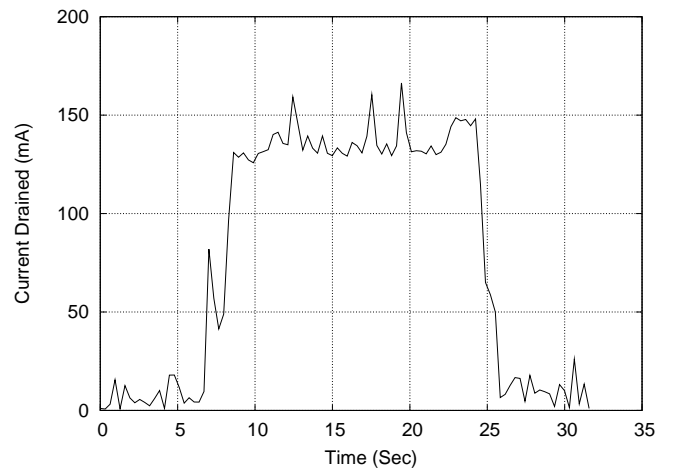

Fig. 2. The power consumption pattern of GPS on an HTC Dream.

### B. Power Profiling Results

We focus on the power consumption of the localization sensors: the GPS, accelerometer, and compass.

*1) GPS Power Consumption:* We studied GPS power consumption by running an application that queries the GPS sensor of the HTC Dream. We did our measurements in both outdoors and indoors environments. Figure 2 shows the typical GPS behavior. When the GPS chip is turned on, it drains a constant current of about 135mA. The GPS chip is turned off as soon as a fix is obtained. The duration required for a GPS fix depends on many parameters, e.g. the surroundings of the device and/or the clearness of the sky. The power pattern has a knee at the start and end of the pattern. This is the result of activating both the GPS chip and the processor.

*2) Accelerometer/Compass Power Consumption:* A similar approach was followed to measure the power consumption of the accelerometer and compass. The HTC Dream has a AK8976A chipset that provides the functionality of **both the accelerometer and compass**. Therefore, the same power is consumed when using the accelerometer, the compass, or both. This has been verified by our experiments. The AK8976A chip is a 6-axis electronic compass that combines a 3-axis geomagnetic sensor with a 3-axis acceleration sensor in an ultra-small package. The Android API provides four modes for querying the accelerometer and the compass that differ in the rate of delivering sensors events. They are called Normal, UI, Game, and Fastest. The rates at which the events are delivered are 4, 10, 20, and 50 events/sec respectively. This is only a hint to the operating system. Events may be received faster or slower than the specified rate [17]. Figure 3 shows the current drained when using each of the data query modes and Table I summarizes the results. As indicated before, the average data rate is a hint to the operating system. Therefore, the actual rate can be higher or lower than requested. This explains why the current drained is not proportional to the data rate. We choose to use the Normal data rate as it fits our need and consumes the least amount of power. Moreover, the operating system is continuously querying the accelerometer sensor with this rate to support the functionality of screen auto-rotation. Consequently using this rate adds zero extra power consumption.

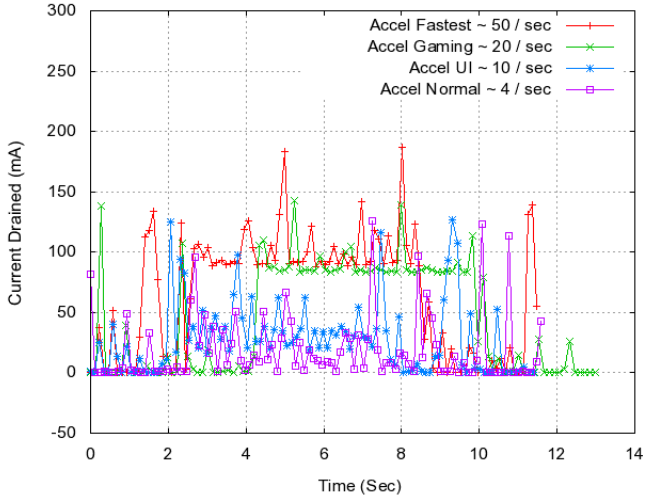

Fig. 3. The power consumption pattern of accelerometer for different data query rates on HTC Dream

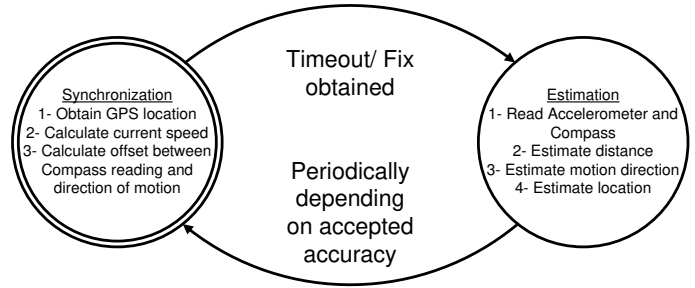

TABLE I
AVERAGE CURRENT DRAINED BY ACCELEROMETER/COMPASS FOR DIFFERENT DATA QUERY MODES ON HTC DREAM. THE AVERAGE DATA RATE IS A HINT TO THE OPERATING SYSTEM. THEREFORE, THE ACTUAL RATE CAN BE HIGHER OR LOWER THAN REQUESTED. THIS EXPLAINS WHY THE CURRENT DRAINED IS NOT PROPORTIONAL TO THE DATA RATE.

| Query Mode | Average Data Rate (Hz) | Average Current Drained (mA) |
|---|---|---|
| Fastest | 50 | 95 |
| Gaming | 20 | 90 |
| UI | 10 | 25 |
| Normal | 4 | 15 |
| GPS | N/A | 135 |

Fig. 4. Finite state machine showing the operation of our system.

### C. Discussion

The power profiling experiment shows that GPS is using the highest amount of power among the available localization sensors. The accelerometer chipset, including the compass too, consumes different amounts of power depending on the rate at which it reports the sensor readings. The accelerometer chip power consumption is not a linear function of the data rate. Our hypothesis is that the chipset has two modes of operation: the high speed query mode and low speed query mode. Unfortunately, we could not get access to the chip data sheet for confirmation. Using the Normal query mode for the accelerometer provides enough functionality for localization and saves more than an order of magnitude in terms of energy if both the GPS and accelerometer run for the same duration.

## III. GAC: THE PROPOSED HYBRID LOCALIZATION SCHEME

In this section, we present the details of our proposed hybrid GPS-accelerometer-compass scheme. We start by an overview of our system followed by the mathematical model and the details of the technique.

### A. System Overview

Figure 4 shows a finite state machine representing our system's operation. The main idea is to depend on the accelerometer and compass to obtain a low-energy localization technique and with a low frequency synchronize with the GPS to reduce the accumulated error.

The location estimator queries the accelerometer and compass periodically to determine the direction of the tracked object and the displacement and speed along this direction. This is combined with the previous location estimate to provide the new location estimate. Periodically, based on a preset value, the GPS is queried to obtain a better location estimate and the current estimate is set to the GPS location. The GPS is also used to determine the initial position and velocity. The frequency of synchronization with the GPS represents a tradeoff between energy-consumption and error in localization. We quantify this tradeoff in Section IV.

### B. Mathematical Model

*1) Assumptions:* Assume that time is discretized into instances at which the accelerometer and compass readings are

queried with a sampling interval $T$. We also assume that the mobile position is fixed relative to the tracked object, e.g. attached to the dashboard of a car. Therefore, its orientation relative to the tacked object's coordinate system is fixed. This is a reasonable assumptions for many applications, e.g. a car navigation system. We also assume that the tracked object moves with a constant direction and acceleration within a query period. Since we use the Normal query mode, which has a 250ms query rate, this is a reasonable assumption.

*2) Approach:* Let $p(n)$ and $v(n)$ be the mobile position (in lat-long coordinate system) and velocity at time instance $n$. Given that $d(n)$ is the vector representing the direction of the mobile movement, obtained by the compass, at time instance $n$ and $a(n)$ is the magnitude of the acceleration in the direction of movement obtained by the accelerometer, the displacement $l$ along the direction of movement is equal to:

$$l(n) = v(n)T + \frac{1}{2}a(n)T^2$$

where $T$ is the sensors sampling interval.

Note that since the tracked object is moving on the surface of the Earth's spheroid, Ecludian distance cannot be used. Instead, we use Vincenty's formula [18] which computes the location of a point which is a given distance and direction from another point on the surface of an oblate spheroid. Vincenty's formula have been widely used in geodesy because it is accurate to within 0.5mm on the Earth ellipsoid. Therefore, the new position of the mobile phone is estimated as:

$p(n+1) = f[p(n), l(n), d(n)]$

Where $f$ is a function that returns the new position of the mobile phone using the direct Vincenty's formula [18] given its current position, direction of movement, and displacement along this direction.

In addition, the new estimated velocity of the phone is given by:

$v(n+1) = v(n) + \int_0^T a(n)dx = v(n) + a(n)T$

To reduce the effect of accumulated error, the system periodically, with duration $TG_{synch}$ synchronizes with the GPS. At this time, the current position and velocity of the mobile phone are set to those obtained from the GPS.

*3) Practical considerations:* Although we assume that the mobile phone is fixed relative to the tracked object, it may happen that the phone orientation slightly changes with relative to the tracked object, e.g. due to hard bumps in roads or a sudden sharp turn. Our approach to compensate for this is that using the GPS synchronization points, we can identify segments in the trip where the speed is constant. While the mobile is moving with constant speed, the acceleration measured by the accelerometer is due to gravity only and can be used to estimate the current orientation of the phone. Knowing the current orientation can be used to correct further accelerometer readings to get better estimate for the acceleration of the mobile device. Speed is detected to be constant during a driving segment if the difference between the maximum and minimum values of the speed during this segment does not exceed specific threshold. Our experiments showed that 0.5 m/s is a suitable value for that threshold.

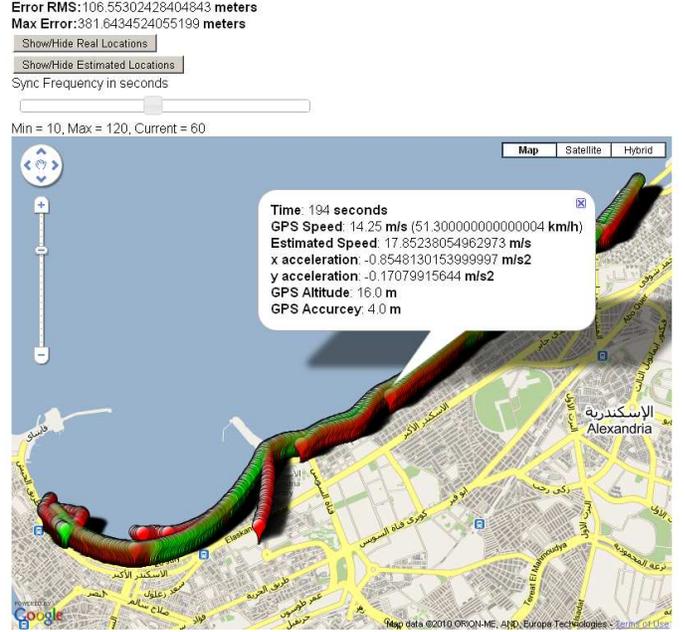

Fig. 5. Snapshot of The Map-based Visual Debugging Tool

| Testbed | Num. of trips | Area covered | Location estimation rate | Duration |
|---|---|---|---|---|
| One (In city) | 4 | 3Km$^2$ | one/sec. | 77 min. |
| Two (Highway) | 4 | 20Km (linear) | one/sec. | 56 min. |

TABLE II
COMPARISON BETWEEN THE TWO TEST ENVIRONMENTS.

In addition, to reduce the noise in the accelerometer/compass readings, we average a number of readings and use this average in the estimation process. Our findings show that a value of four gives good accuracy with minimum effect on latency.

## IV. PERFORMANCE EVALUATION

In this section, we present the evaluation of our scheme compared to the Simple Linear Predictor presented in EnLoc [19]. The Simple Linear Predictor of EnLoc suggests that the location of the phone at some time instant can be estimated as the linear extrapolation of its previous two locations estimates. They suggested that this is effective when moving on straight lines. So in essence, our approach is similar to EnLoc except that we use the accelerometer to obtain a better estimate of location between synchronization points.

We start by describing our experiment setup followed by the evaluation results.

### A. Experiment Setup

In order to evaluate the proposed scheme, we used an HTC Dream phone fixed to the dashboard of a car and developed an application to log the readings of GPS, accelerometer, and compass sensors. We drove in two different environments. The first set of traces were collected in Alexandria which represents a typical large city with slow traffic and large number of turns. The second set of traces are for driving on the highway, which is the best environment for EnLoc.

We have different driving patterns including constant and variable speed, straight and curved roads, and roads with sharp turns. Table II summarizes the characteristics of both environments.

We have also developed an interactive map-based debugging tool that helped in tweaking the scheme and visualizing the impact of the rate of synchronization with GPS on the accuracy of the estimated locations. A snapshot of this tool is shown in Figure 5. In all experiments GPS was assumed to be the global reference with zero location estimation error.

### B. Energy Savings

Figure 6 shows the power consumption of the proposed scheme with increasing GPS synchronization period $TG_{synch}$. The figure shows that both GAC and EnLoc reduce energy considerably compared to the GPS. In the figure, GAC results do not take into account the fact that by using the Normal query mode of the accelerometer, our scheme does not incur any additional power consumption as this mode is used continuously by the mobile phone to detect tilting of the mobile. If this is taken into account, as shown by the GAC-AccFree results in Figure 6, the power consumption of our proposed scheme and EnLoc are the same.

In addition, the figure shows that as the GPS synchronization period increases, the energy savings exponentially increases. This shows the potential of our proposed technique to considerably save the battery of the phone.

### C. System Accuracy

We used localization Root Mean Square of Error (RMSE) to evaluate the accuracy of the different schemes. The RMSE is defined as:

$$RMSE(TG_{sync}) = \sqrt{\sum_{i=1}^{N} \frac{H[L_{est}(t_i, TG_{sync}), L_{true}(t_i)]^2}{N}}$$

Where:
- $L_{true}(t_i)$: the true location, calculated by GPS at time $t_i$.
- $L_{est}(t_i, TG_{sync})$: the location estimated by our scheme with period between GPS synchronization points equals to $TG_{sync}$ at time $t_i$.
- $N$: number of location estimates during the $TG_{sync}$ period. Note that $N = \lfloor TG_{sync}/T \rfloor$.
- $H[L_{est}(TG_{sync}) - L_{true}]$ is the Haversine distance [20] between the estimated location and the true location. Haversine distance gives the great-circle distance between two points on a sphere from their longitudes and latitudes. Again, Ecludian distance cannot be used due to the curvature of the Earth's surface.

Figure 7 shows the performance of our system on highways, which is the perfect environment for EnLoc. The figure shows that our scheme performance is comparable to EnLoc's. For higher values of synchronization period, the EnLoc scheme outperforms our scheme because of inaccuracies of measurements of sensors and accumulation of error. However, for synchronization periods less that 60 seconds, the energy saving is about one order of magnitude for both techniques with no significance difference in accuracy between them.

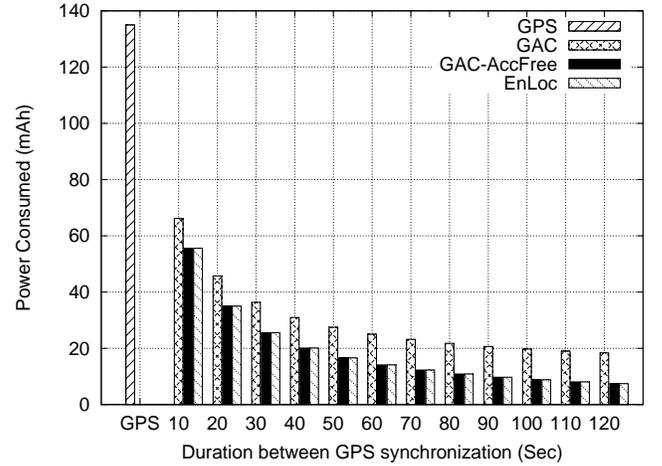

Fig. 6. The total power consumed during an hour by the proposed GAC scheme for different values for duration between synchronization points $TG_{synch}$. The first bar represents the GPS power consumption for comparison.

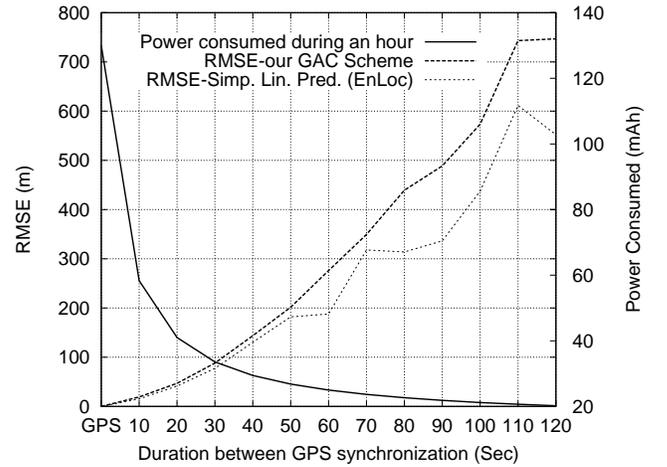

Fig. 7. Performance of our GAC scheme vs EnLoc on highways.

The **main advantage** of our scheme comes inside cities as shown by Figure 8. The figure shows that our scheme can significantly outperform the EnLoc scheme even for short synchronization periods. The benefit obtained by our scheme increases as the synchronization period increases. The movement in straight lines assumptions of EnLoc is clearly violated for intra-city driving.

Figures 7 and 8 also show the power savings compared to the GPS (Figure 6 replotted on both figures). The figure shows that the proposed technique leads to exponential saving in energy with a linear degradation in accuracy.

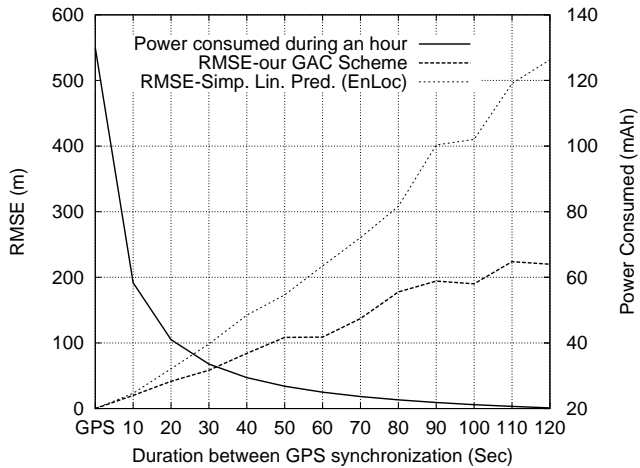

Fig. 8. Performance of our GAC scheme vs EnLoc inside the city.

## V. Conclusion and Future Work

In this paper, we proposed GAC as an energy-efficient localization scheme that exploits the accelerometer and the compass existing in many modern mobile devices nowadays. We started by presenting an energy profile for the GPS, accelerometer, and compass on an Android-based HTC Dream phone. Our profiling results show that the phone runs the accelerometer and compass simultaneously, as they are provided by the same chip. Moreover, the energy consumption of the accelerometer-compass chip is not linear in the query rate, as the required query rate is only a hint to the operating system. In addition, GPS consumes about an order of magnitude more energy than the combined accelerometer-compass when run for the same time. Our GAC scheme exploit this fact to provide an energy efficient localization system for mobile phones that depends mainly on the accelerometer-compass and switches to the more energy-consuming GPS only infrequently for synchronization. We also discussed how our scheme handles shaking and minor changes in orientation along with reducing noise in sensors measurements.

Our experiments showed that our scheme provides exponential saving in energy compared to GPS with linear loss in accuracy. In addition, compared to a linear prediction scheme, it gives significant accuracy gain (up to 97%) within intra-city roads with comparable performance on highways.

Currently, we are working in extending our work in multiple directions including applying GAC to indoor localization, using a Kalman filter to further handle errors in measurements, dynamically changing the synchronization period based on error estimation, among others.

## Acknowledgment

This work is supported in part by a Google Research Award.